\documentstyle[sprocl,psfig]{article}

\def\Journal#1#2#3#4{{#1} {\bf #2}, #3 (#4)}

\def\be{\begin{equation}}
\def\ee{\end{equation}}
\def\bea{\begin{eqnarray}}
\def\eea{\end{eqnarray}}
\begin{document}

\title{
HADRONIZATION IN HEAVY-ION REACTIONS
}

\author{T. S. BIR\'O\footnote{Talk held at CF98 Workshop, M\'atrah\'aza, Hungary}, P. L\'EVAI,  J. ZIM\'ANYI}

\address{Research Institute for Particle and Nuclear Physics \\
H-1525 Budapest P.O.Box 49, Hungary\\E-mail: tsbiro@sunserv.kfki.hu} 

\author{ AND \\}

\address{}

\author{ C.  T. TRAXLER}

\address{Institute f\"ur Theoretische Physik \\
D-35392 Giessen, Heinrich-Buff-Ring 15, Germany \\
E-mail: Chris.Traxler@theo.physik.uni-giessen.de}


\maketitle\abstracts{We present a model of fast hadronization
of constituent quark matter in relativistic heavy ion 
collisions based on rate equations and capture cross sections
in non-relativistic potential. We utilize a thermodynamically
consistent approach with a non-ideal equation of state
including correlation terms based on string phenomenology.
We investigate strange and non-strange particle ratios
observed in CERN SPS experiments. }

\section{Correlations, Clusters, Confinement}

Quark matter in the moment of hadronization is one of
the most strongly correlated system we deal with in
physics. Confinement forces are long range forces,
so no "ionization" of bound clusters of constituent quarks
is possible below the deconfinement temperature.
In order to work with an "effective" theory of the
strongly interacting quark matter, even in its simplest
form of constructing an descriptive phenomenological
equation of state, we have to take into account effects
of correlation.

%Fig.\ref{trans1} shows a 
A comparison with well known hard core problem
in nuclear matter may be enlightening. 
Although the underlying pair potential
diverges at zero distance ($r \rightarrow 0$), the relative
wave function is zero at this point due to exactly this
repulsive core. As a consequence the effective pair potential
is finite at small distance giving rise to finite corrections
to the mean field energy (correlation energy).
In case of the confining static potential between massive
quarks the relative wave function vanishes beyond a
characteristic distance (the Bohr radius of hadronic bound state).
Here also the divergence at $r \rightarrow \infty$ is regularized 
by the pair correlations and the modification to the mean field
energy and pressure is finite.

The above consideration results in an effective equation of state 
at the characteristic particle density determined by the
minimum of the effective pair potential. This concept can be, however,
extended to all densities: the correlation improved pair potential
results in a better high-density behavior of nuclear and in
a better low-density behavior of quark matter.
Eventually a certain, nonlinear density-dependence in the
equation of state takes such correlations into account
in a phenomenological way.

%%%%%%%%%%%%%% FIG 1 %%%%%%%%%%%%%%%%%%

%\begin{figure}[t]
%\rule{5cm}{0.2mm}\hfill\rule{5cm}{0.2mm}

%\vskip1.0cm
%\centerline{\psfig{figure=fig1.ps,height=5.0cm,width=4.0cm}}

%\rule{5cm}{0.2mm}\hfill\rule{5cm}{0.2mm}
%\caption{The qualitative shape of the  pair potential, the correlation functions
%and the folded effective potential, respectively.
%The upper line shows the nuclear hard core effect, while the
%lower line of pictures the effect of confinement of quarks
%into hadrons. \label{trans1}}
%\end{figure}

%%%%%%%%%%%%%%%%%%%%%%%%%%%%%%%%%%%%%%%%%%

In our present model massive quarks and their hadronic 
(i.e. color neutral) clusters are constituents of a mixture.
We describe this mixture with a quasi particle Hamiltonian
including a density dependent background energy density,
\be
{\cal H} = \sum_i \left( \omega_i(k) + U_i(n) \right) \, n_i + U(n).
\ee
The self-consistency of the definition of the chemical potential,
\be
\mu_i = \frac{\partial {\cal H}}{\partial n_i} = 
\omega_i + U_i,
\ee
requires
\be
\sum_j n_j \frac{\partial U_j}{\partial n_i} + 
\frac{\partial U}{\partial n_i} = 0,
\ee
and the variational minimum with respect to the temperature leads to
\be
\frac{\partial {\cal H}}{\partial T} =
\sum_j n_j\frac{\partial U_j}{\partial T} +
\frac{\partial U}{\partial T}.
\ee
These thermodynamical consistency relations are satisfied by the
following general form of the free energy
\be
 F(N_1, N_2, \ldots, V, T) = \sum_i F^{{\rm id}}(N_i,V,T)
 + V \,   b(N_1/V, N_2/V, \ldots).
\label{FREE}
\ee
Here $N_i = Vn_i$ is the number of constituents, $V$ the reaction
volume and $T$ the temperature. We choose the free energy
as the thermodynamical potential, because for a system of
"quark to hadron transition" - chemical, shortly
"transchemical", equations the time evolution of the numbers
can be obtained relatively easy, contrary to that of the
chemical potentials $\mu_i$.
Note that the $b(n_1,n_2,\ldots)$ correction does not depend on the temperature
or on the volume explicitly. The derived thermodynamical quantities are
\bea
 \mu_i &=& \mu_i^{{\rm id}} + \frac{\partial b}{\partial n_i},
\nonumber \\
 S &=& S^{{\rm id}},
\nonumber \\
 E &=& E^{{\rm id}} + V b.
\nonumber \\
 p &=& p^{{\rm id}} - b + \sum_i n_i \frac{\partial b}{\partial n_i}.
\label{THERMO}
\eea
The relation to the vector mean field $U_i$ and to the background
(bag) energy density is given by
\bea
U_i &=& \frac{\partial b}{\partial n_i}, \nonumber \\
U   &=&  b - \sum_i n_i \frac{\partial b}{\partial n_i}.
\eea
Let us consider a few examples. For the good old MIT Bag model
$b=B$ is constant, leading to $U=B$ and vanishing vector mean
field for the constituents $U_i=0$. The pressure
$p = n_qT - B = aT^4-B$ is negative below a critical $T=T_c$,
where the quark matter is unstable.

In a pure mean field approach $b$ is linear,
$b = \sum_i B_i n_i,$ leading to $U=0$ (no bag constant) and
$U_i = B_i$. This system is stable at all temperatures ($p=p^{{\rm id}}$).

Finally we consider a simple system of massive quarks interacting
via a string-like potential. In this case the correction to the
free energy is proportional to the average distance of quarks,
\be
b = \, \kappa \, n \, \langle \ell \rangle \,  = \, \sigma n^{2/3}.
\ee
In this case one obtains corrections to both mean fields,
$U=(1/3)\sigma n^{2/3}$ and $U_q = (2/3) \sigma n^{2/3}$.
The correlation correction is distributed between the background
("bag" constant) and the vector mean field for the quarks.
The pressure of this system also becomes negative below
a certain temperature:
\be
p = nT - \frac{1}{3} \sigma n^{2/3} \approx
aT^4 - bT^2.
\ee
This property is common with the bag model. The string EOS, however,
results zero pressure at zero density (so no vacuum renormalization
is needed). Furthermore - due to the vector mean field correction -
the string EOS shows a non-ideal behavior of the chemistry, which
is absent in the bag model.

\begin{figure}[h]
\rule{5cm}{0.2mm}\hfill\rule{5cm}{0.2mm}

\vskip1.0cm
\centerline{\psfig{figure=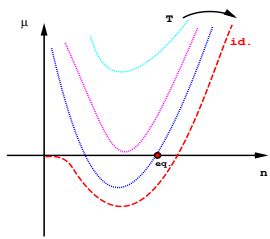,height=3.0cm,width=3.5cm}}

\rule{5cm}{0.2mm}\hfill\rule{5cm}{0.2mm}
\caption{Isotherms of the chemical potential are plotted against
the quark density with a non-ideal equation of state
with string-like interaction. The infinite temperature case
coincides with the ideal gas form, while below a critical
temperature there is no chemical equilibrium possible:
the quark density tends towards zero. \label{chem}}
\end{figure}

Fig.\ref{chem} shows the dependence of the quark chemical potential on
the density,
\be
\mu = d \, \, \ln \frac{n}{cT^3} + \frac{2}{3} \sigma n^{-1/3}.
\ee
Since the entropy production in an isolated expanding system
is given by
\be
\dot{S} = - \sum_i \frac{\mu_i}{T} \dot{N}_i \ge 0,
\ee
any physical description of the coalescence rate of quarks
possesses the property that $\mu$ and $\dot{N}$ have opposite signs.
In the particular two-body fusion process studied in the transchemistry
model of next Section, one has
\be
\dot{N} \propto -  N^2 \left( 1 - e^{-\mu/T} \right),
\ee
satisfying the above condition.

Due to this often cited entropy growth principle, the quark density
in a constant volume
decreases whenever $\mu > 0$ and increases whenever $\mu < 0$.
The isotherms of the $\mu(n)$ relation depicted in Fig.\ref{chem} reveal
that below a given temperature $T_{{\rm chem}}$ no chemical
equilibrium is possible in a pure quark matter with string EOS.
In this case the quark chemical potential is overall positive
at any density --- so quarks will be eliminated from the system.
The only natural way to find a thermodynamically stable state
is to form color neutral clusters which makes no contribution
to the string correction in the equation of state.
This is what happens by the hadronization of colored constituent
quark matter and this is the basis of the transchemistry
model discussed in the next section.

%%%%%%%%%%%%%% FIG 2 %%%%%%%%%%%%%%%%%%

%%%%%%%%%%%%%%%%%%%%%%%%%%%%%%%%%%%%%%%%%%%%%%%%%%%%%%%%%

\begin{figure}[t]
\rule{5cm}{0.2mm}\hfill\rule{5cm}{0.2mm}

\vskip0.5cm
\centerline{\psfig{figure=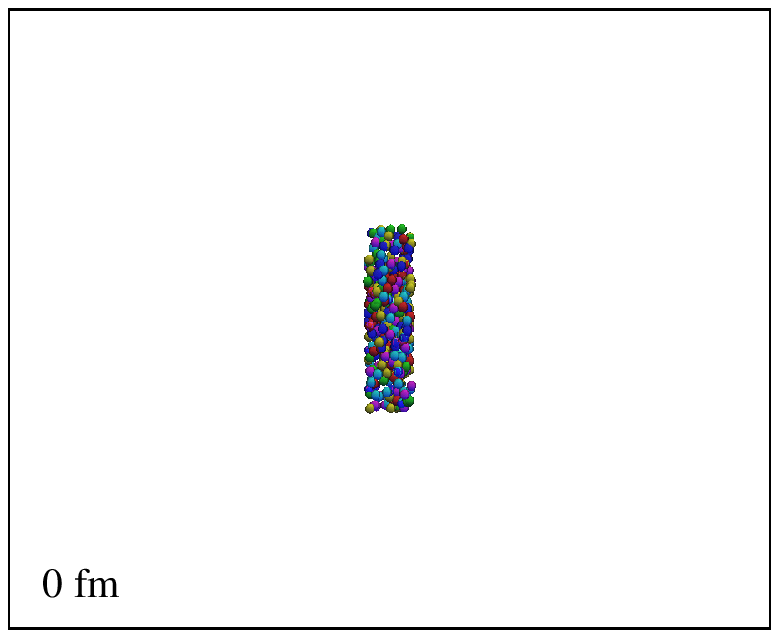,width=38mm,height=44mm}
            \psfig{figure=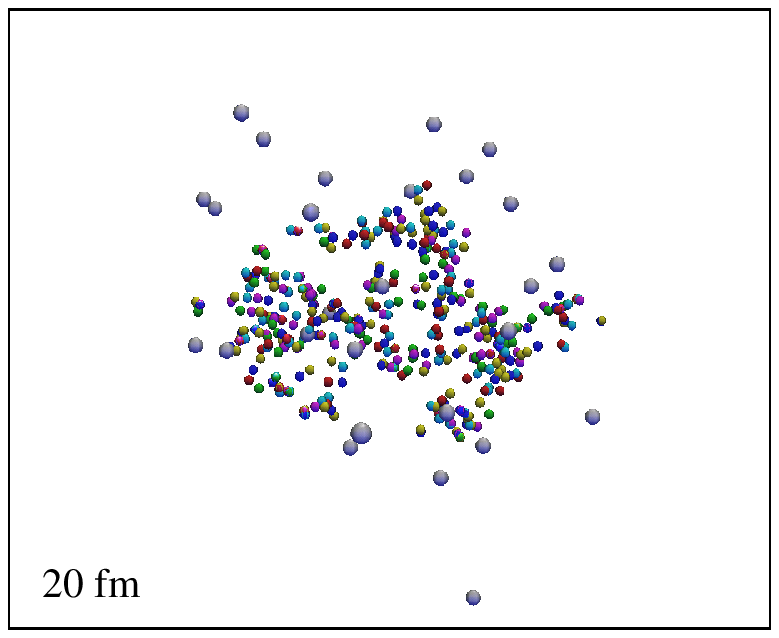,width=38mm,height=44mm}
            \psfig{figure=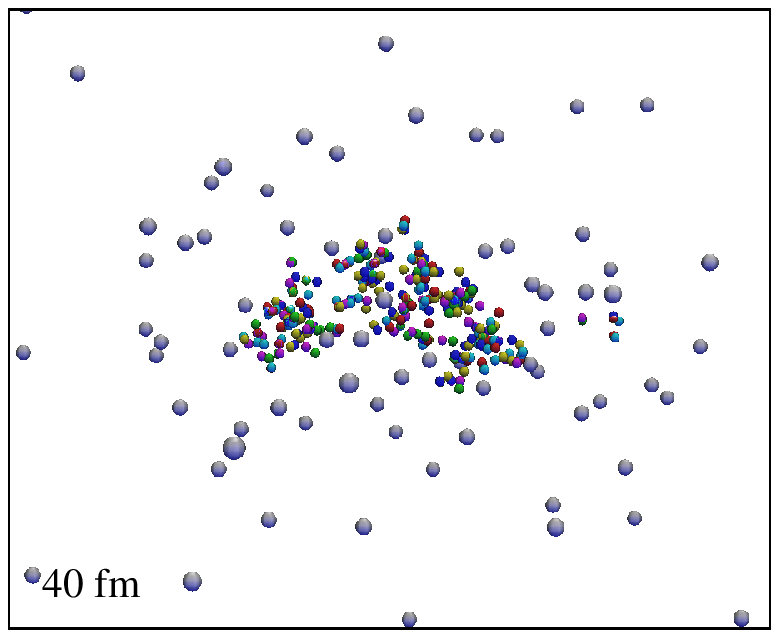,width=38mm,height=44mm}}

\rule{5cm}{0.2mm}\hfill\rule{5cm}{0.2mm}
\caption{\em Simulation snapshots after initialization of
a central-rapidity slice of a QGP tube with Lorentz-invariant flow
(``Bjorken scenario'') in the chromodielectric model. 
Colored objects (quarks, gluons) are drawn in dark grey,
hadrons in light grey.}
\label{bjorken}

\end{figure}

Finally we would like to show some results of a molecular
dynamical calculation of massive quark matter in the framework
of the chromodielectric model. This model describes the
confining strings due to an abelian Gauss law but with
a nontrivial dielectric constant.  Inspecting the snapshots of
quark matter evolution in a Bjorken scenario (Fig.\ref{bjorken})
the ramification is clearly seen in the intermediate stage
of hadronization.

%%%%%%%%%%%%%%% FIG 2B %%%%%%%%%%%%%%%%%%%%

%%%%%%%%%%%%%%%%%%%%%%%%%%%%%%%%%%%%%%%%%%%%%%%%%%%%%%%%%

\section{The transchemistry model}

A detailed description of the transchemistry model can be found
in \cite{Trchem}. The equation of state used for the description
of the evolution of constituent quark matter into hadrons
during a sudden expansion utilizes a generalization of the
above discussed string-like correlation energy.
In the hadronizing mixture we deal with quarks, anti-quarks,
diquarks, anti-diquarks carrying color charge and with mesons,
baryons and anti-baryons being color neutral. The number of strings
is assumed to be proportional to a weighted sum of color particles
$Q = \sum_i q_i N_i.$ We take $q_i=0$ for hadrons, $q_i=1$ for
quarks and $q_i=3/2$ for diquarks. The average length,
$\langle \ell \rangle = n_c^{-1/3}$ is determined by the color
density $n_c = \sum_{i=c} N_i/V$ excluding hadrons from the sum.
The correction to the free energy of a mixture of massive ideal gas
constituents without inner degrees of freedom is then given by
\be
\Delta F = \sigma n_c^{-1/3}Q.
\ee
The modification of the non-hadronic chemical potentials becomes
\be
\mu_i = \mu_i^{{\rm id}} + \sigma n_c^{-1/3}
\left( q_i - \frac{1}{3} \frac{Q}{Vn_c} \right).
\ee
The non-ideal cooling law, equivalent to $\partial_{\mu} T^{\mu \nu} = 0$
for a perfect fluid of the mixture, becomes
\bea
\frac{\dot{T}}{T} &=& - \frac{2}{3} \frac{\dot{V}}{V}
- \frac{\sum_i \dot{N}_i}{\sum_i N_i}
- \frac{2}{3} \frac{\sum_i (m_i/T)\dot{N}_i}{\sum_i N_i}
\nonumber \\
&-& \frac{2}{3} \frac{\sum_i (\Delta\mu_i/T)\dot{N}_i}{\sum_i N_i},
\eea
with the additional term
\be
\sum_i (\Delta\mu_i/T)\dot{N}_i = \sigma \sum_{i=c}
n_c^{-1/3} \left( q_i - \frac{1}{3} \frac{Q}{Vn_c} \right)
\dot{N}_i,
\ee
re-heating the system while eliminating color charges.
The cooling is due to the expansion ($\dot{V} > 0$) and due to
rest mass production by making hadronic resonances.

The initial temperature and volume was obtained in a 
one dimensional Bjorken scenario from the bombarding energy
and the stopping power. At the CERN SPS $Pb+Pb$ experiment
we obtain $V_0 \approx 400$ fm$^3$ and $T_0 \approx 180$ MeV.
The initial quark numbers include direct and
gluon fragmentation pair creation and sum up to about
$3000$ with a strangeness ratio of $f_s = 0.21$.

The general two body fusion process is supported by
the in-medium cross section of massive constituents
in a Coulomb potential in medium. This is proportional
with a screening volume, $\rho^3$, which also depends on
the presence of color charges. We make the assumption that
$\rho^3 n_c$ is constant obtaining a dynamical confinement
effect of increasing hadronization cross sections in the
low color density stage. These phenomenologically amounts to
the string-like correlation correction to the hadronization
cross sections and influences the dynamics of the transchemical
evolution.

Each chemical reaction is of type $i + j \leftrightarrow k$,
accounting for the changes $dN_i=dN_j = -Adt$ and
$dN_k = Adt$ in the particle numbers. The general form,
\be
A = R_{ij \rightarrow k} N_i N_j 
\left( 1 - e^{\frac{\mu_k}{T} - \frac{\mu_i}{T} - \frac{\mu_j}{T} }
\right),
\ee
assures the increase of entropy while approaching chemical equilibrium.
Finally, in order to get the final hadron numbers, hadronic decays
are taken into account with the dominant branching ratios obtained
from Particle Data Table\cite{PDT}.

\section{Constituent Quark Matter at SPS}

\begin{figure}[htb]
\rule{5cm}{0.2mm}\hfill\rule{5cm}{0.2mm}

\centerline{\psfig{figure=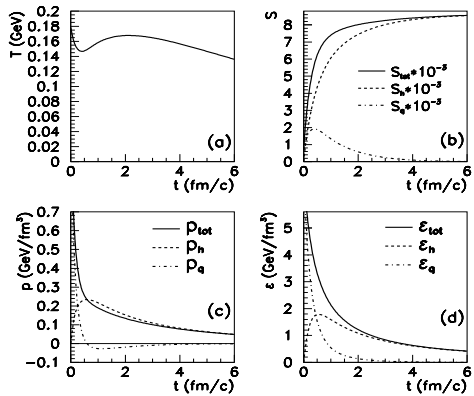,height=6.0cm,width=7.0cm}}

\rule{5cm}{0.2mm}\hfill\rule{5cm}{0.2mm}
\caption{Evolution of thermodynamical quantities
in the transchemistry model \label{qmfig1}}
\end{figure}

In this section we present results of the numerical solution
of 43 coupled differential equations describing the chemical
and thermal evolution of the quark, diquark - hadron mixture.
In Fig.\ref{qmfig1} the evolution of the temperature, entropy,
pressure and energy density is shown for a 168 GeV/nucleon
$Pb+Pb$ collision. At the beginning there is a rapid drop of
the temperature (cf. Fig.\ref{qmfig1}a)
due to heavy hadron resonance formation.
It is followed by a mild re-heating as an effect of the
color confinement (by producing a colorless hadron the
kinetic energy of the relative motion and the string energy
is removed and replaced by the rest mass of the hadron).
The total entropy monotonically increases during the hadronization
while the contribution of colored particles is gradually
eliminated (Fig.\ref{qmfig1}b).
The partial pressure of the constituent quark plasma rapidly
decreases, it becomes even negative as the color density
drops (Fig.\ref{qmfig1}c). The total pressure, however,
remains positive - so the mixture is stable and continues
the adiabatic expansion.
The partial and total energy densities  evolve in the 
expected way going over to the $\epsilon \propto 1/\tau$
behavior at very late times only, when $p=0$.

%%%%%%%%%%%%%%% FIG 3 %%%%%%%%%%%%%%%%%%%%

%%%%%%%%%%%%%%%%%%%%%%%%%%%%%%%%%%%%%%%%%%%%%%%%%%%%%%%%%

\begin{figure}[htb]
\rule{5cm}{0.2mm}\hfill\rule{5cm}{0.2mm}

\centerline{\psfig{figure=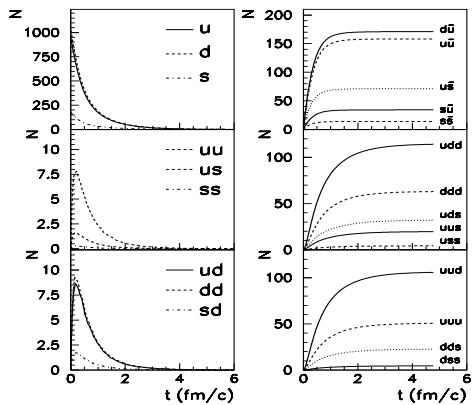,height=6.0cm,width=7.0cm}}

\rule{5cm}{0.2mm}\hfill\rule{5cm}{0.2mm}
\caption{Evolution of thermodynamical quantities
in the transchemistry model \label{qmfig23}}
\end{figure}

Fig.\ref{qmfig23} presents the time evolution of
constituent particle numbers: quarks, diquarks
(left column from top to bottom), mesons and
baryons (right column). The evolution of anti-particles
is qualitatively similar, their absolute numbers are
only smaller. The hadron numbers shown here are those 
before hadronic resonance decays.
It is clear from these figures that mesons are formed
faster than baryons, the reason being that in this
model the baryons are produced by a two step process
and the initial diquark content was zero in these
particular calculations.

%%%%%%%%%%%%%%% TABLE TEMPLATE %%%%%%%%%%%%
 \begin{table}[htb]

 \caption{ hadron multiplicities for $Pb+Pb$ collision
  at 168 GeV/nucleon bombarding energy. \label{tab:exp}}
 \vspace{0.2cm}

 \begin{center}
 \footnotesize
 \begin{tabular}{|c|c|c|c|c|}
 \hline

 {Pb+Pb} & NA49 & TrCHEM. & ALCOR & RQMD \\

 \hline

  $K^+$  &  76 & 76.14 & 78.06 & 79.0 \\

  $K^-$  &  32 & 38.36 & 34.66 & 50.4  \\

  $K^0_S$ & 54 & 57.25 & 56.36 & 63.5 \\

 $p-\bar{p}$ & 145 & 151.6 & 147.03 & 171.8 \\

 $\Lambda^0$-like & 50$\pm$ 10 & 62.30 & 69.07 & 56.8 \\

 $\bar{\Lambda}^0$-like & 8 $\pm$ 1.5 & 8.14 & 8.12 & 19.3 \\

 \hline

\end{tabular}
\end{center}
\end{table}

%%%%%%%%%%%%%%%%%%%%%%%%%%%%%%%%%%%%%%%%%%%%%%%%

In the ALCOR model \cite{ALCOR} the ratio of hadronic species
are determined by the steepness of these curves
(i.e. by $\dot{N}_i(0)$). Since the $N_i(t)$ curves do not
cross, the algebraic approach of ALCOR to the solution of
the underlying rate equations works well.

%%%%%%%%%%%%%%% FIG 4 %%%%%%%%%%%%%%%%%%%%

%%%%%%%%%%%%%%%%%%%%%%%%%%%%%%%%%%%%%%%%%%%%%%%%%%%%%%%%%

Finally in Table \ref{tab:exp} some hadron numbers obtained
in the transchemistry model are compared with experimental data
of the NA49 collaboration and with two other theoretical
models: ALCOR and RQMD\cite{RQMD}. A more detailed comparison
can be found in \cite{Trchem}.

\section{Conclusion}

We presented a new model for the hadronization of constituent
quark plasma based on a string-like correlation energy in the
equation of state and on rate equations in a quark matter -
hadron matter mixture. Due to large hadronization rates
and due to a very off-equilibrium initial state, over-saturated
with (massive) quarks, the hadronization process is fast:
its time-scale is a few fm/c. We also found that in the
particular situation of 168 GeV/nucleon heavy-ion collisions
the algebraic ALCOR model is a good approximation to the
numerical solution of the transchemical rate equations.
The comparison with presently available experimental data
indicate, that in such collisions a piece of matter is
formed, inside which massive quarks, diquarks and their
anti particles interact with a string-like mean field.

\section*{Acknowledgments}

This work was supported by the Hungarian National Science
Research Fund OTKA (T024094, T019700) and by a common
project of the Deutsche Forschungsgemeinschaft and
the Hungarian Academy of Science (DFG-MTA 101).

%%%%%%%%%%%%%%%%%%% REFS %%%%%%%%%%%%%%%%%%%%%%%%%%%%%%%%%

\end{document}